\documentclass[11pt,sort&compress]{elsarticle}

\input{preamble.tex}

\begin{document}

\hypersetup{
  linkcolor=darkrust,
  citecolor=seagreen,
  urlcolor=darkrust,
  pdfauthor=author,
}

\begin{frontmatter}
    \title{Bayesian optimal design accelerates discovery \\ of soft material properties from bubble dynamics}
    
    \author[add1]{Tianyi Chu\corref{cor1}}
    \ead{tchu72@gatech.edu}
    \author[add2]{Jonathan B.\ Estrada}
    \author[add1,add3,add4]{Spencer H.\ Bryngelson}
 
    \address[add1]{School of Computational Science $\&$ Engineering, Georgia Institute of Technology, Atlanta, GA 30332, USA\vspace{-0.125cm}}
    \address[add2]{Department of Mechanical Engineering, University of Michigan, Ann Arbor, MI 48105, USA\vspace{-0.125cm}}
    \address[add3]{Daniel Guggenheim School of Aerospace Engineering, Georgia Institute of Technology, Atlanta, GA 30332, USA\vspace{-0.125cm}}
    \address[add4]{George W.\ Woodruff School of Aerospace Engineering, Georgia Institute of Technology, Atlanta, GA 30332, USA}
    \cortext[cor1]{Corresponding author}
    \date{}
\end{frontmatter}

\begin{abstract}
An optimal sequential experimental design approach is developed to computationally characterize soft material properties at the high strain rates associated with bubble cavitation. 
The approach involves optimal design and model inference.
The optimal design strategy maximizes the expected information gain in a Bayesian statistical setting to design experiments that provide the most informative cavitation data about unknown soft material properties. 
We infer constitutive models by characterizing the associated viscoelastic properties from measurements via a hybrid ensemble-based 4D-Var method (En4D-Var). 
The inertial microcavitation-based high strain-rate rheometry (IMR) method (\citep{estrada2018high}) simulates the bubble dynamics under laser-induced cavitation. 
We use experimental measurements to create synthetic data representing the viscoelastic behavior of stiff and soft polyacrylamide hydrogels under realistic uncertainties.
The synthetic data are seeded with larger errors than state-of-the-art measurements yet matches known material properties, reaching $1\%$ relative error within 10 sequential designs (experiments).
We discern between two seemingly equally plausible constitutive models, Neo-Hookean Kelvin--Voigt and quadratic Kelvin--Voigt, with a probability of correctness larger than $99\%$ in the same number of experiments.
This strategy discovers soft material properties, including discriminating between constitutive models and discerning their parameters, using only a few experiments.

\end{abstract}

\blfootnote{Code available at \url{https://github.com/InertialMicrocavitationRheometry/IMR_Bayesian_design}}

\section{Introduction}

Large and rapid deformations in compliant soft materials, such as those caused by shock waves or lasers, can lead to mechanical failure.
Cavitation may occur when these materials are exposed to tensile waves, leading to high strain rates ($10^3$--$10^8$~\unit{\per\second}).
Energy-focused cavitation, when used appropriately, can benefit biologic, medical, and surgical applications, including tissue phantom studies, laser surgery, and DNA manipulation in target cells~\citep{mancia2019modeling,vlaisavljevich2016visualizing,brujan2006stress,bailey2003physical,brennen2015cavitation}.
However, accurate characterization of realistic soft materials and biotissues under such high strain rates and large deformations is challenging due to their common high compliance~\citep{arora1999compliance, chen2010split}, and mechanical behavior beyond the linear elastic regime~\citep{lin2009spherical,style2013surface}.
Therefore, a faithful representation of the constitutive response of the underlying tissue is required to predict mechanical behavior at high strain rates.

Inertial microcavitation-based high strain-rate rheometry (IMR) has been proposed by \citet{estrada2018high} for characterizing compliant materials at finite deformations and fast speeds. 
This high-strain rate rheometer combines laser-induced cavitation with physical bubble dynamics models to estimate the viscoelastic properties of hydrogels through observations of the bubble radius time history.
The IMR method has been applied to characterize the mechanical behavior of commonly used biomimetic hydrogels, including
polyacrylamide (PA)~\citep{estrada2018high,yang2020extracting,buyukozturk2022particle},
agarose~\citep{mancia2021acoustic,yang2022mechanical}, and gelatin~\citep{bremer2024ballistic}. 
The time efficiency of the cavitation experiments, however, is limited by factors such as the chemical, degassing, and swelling protocols necessary to create pristine samples for characterization~\citep{lopez2014three,estrada2018high}. 
Therefore, an experimental design strategy is necessary to efficiently probe material responses to different physical mechanisms, such as deformation, pressure, and thermal effects, while preserving experimental or computational resources. 
This design approach is intended to be robust for characterizing soft materials under different sources of uncertainty, including variations in experimental configuration and observational noise.
We use the computationally efficient IMR method to develop a simulation-based optimal experimental design (OED) approach for material parameter characterizations and the physical models and theory that underpin them.

The IMR-based OED seeks to optimize the design of cavitation experiments to yield the most informative data about the viscoelastic properties of the unknown material.
Following the decision-theoretic approach by \citet{lindley1956measure}, the relative entropy, or Kullback--Leibler (KL) divergence, from the posterior to the prior within the Bayesian statistical setting is often used to measure the information provided by an experiment. 
Therefore, the design process focuses on optimizing the expectation of this utility function, also known as the expected information gain (EIG).
However, the direct calculation of the EIG is hindered by the intractability of the inherent double-loop integral due to the absence of closed forms and the inability of conventional Monte Carlo (MC) methods.
Nonlinear models complicate the analytical integration of likelihood functions or posterior distributions, necessitating computational methods.
Different approaches have been proposed to numerically evaluate the EIG, including nested Laplace approximations~\citep{lewi2009sequential,cavagnaro2010adaptive,long2013fast,ryan2014towards} and 
nested Monte Carlo (NMC) estimators~\citep{hamada2001finding, ryan2003estimating,huan2013simulation,myung2013tutorial,du2024efficient}.
The Laplace approach systematically introduces bias, though NMC provides accurate estimators using a finite number of Monte Carlo samples. 

Variational methods have also been incorporated into the EIG estimators to improve the convergence rate and accuracy~\citep{foster2019variational,foster2020unified}. 
Readers are referred to \citet{ryan2016review} and \citet{rainforth2024modern} for reviews on this topic. 
With appropriate EIG estimators, the remaining task of Bayesian OED (BOED) is to optimize the EIG within the domain of design variables.
Multiple optimization methods have been considered, such as simulated annealing~\citep{muller2005simulation}, interacting
particle systems~\citep{amzal2006bayesian}, stochastic optimization~\citep{huan2013simulation,huan2014gradient,carlon2020nesterov, karimi2021optimal}, and Bayesian optimization (BO)~\citep{foster2019variational,kleinegesse2020bayesian,hase2021gryffin}.
In this work, BO is the optimizer selected for its data efficiency, robustness to multi-modality, and ability to deal with noisy observations.
We refer the reader to \citet{shahriari2015taking} and \citet{snoek2012practical} for a comprehensive review and practical implementation of BO. 
Instead of using the same design throughout the experimental process, sequential or adaptive designs have gained popularity in Bayesian design literature due to their flexibility and efficiency~\citep{muller2007simulation, myung2013tutorial,kim2014hierarchical,huan2016sequential}.
Unlike fixed experimental configurations in static designs, sequential designs aim to maximize the expected utility at each stage of experimentation based on the outcomes of previous experiments and the possible predictions of future ones.
For cavitation rheometry, inferring material parameters from bubble dynamics data are important for proceeding with the sequential design.

We use data assimilation (DA) techniques for bubble-dynamics-based rheometry to improve predictions in uncertainty-prone high-strain-rate regimes.
We combine the IMR method with observational data such as bubble-radius trajectories.
The information needed to describe complex systems comes from different sources and has different characteristics, such as modeling assumptions and measurement noise.
Each source is unlikely to fully observe the system, leading to information discrepancies between the theoretical model and the data.
DA rectifies this problem by addressing uncertainty in the model and the data.
In particular, the ensemble Kalman filter (EnKF) is an often-used DA tool due to its simple conceptual formulation and relative ease of implementation~\citep{evensen1994sequential}. 
It achieves relatively high accuracy for a small ensemble, approximating the state as a multivariate Gaussian. 
Applications of EnKF include oceanography~\citep{evensen2003ensemble,sakov2012topaz4}, atmospheric science~\citep{houtekamer1998data, burgers1998analysis, whitaker2002ensemble}, and engineering~\citep{aanonsen2009ensemble}.
Other variants of EnKF, such as ensemble Kalman
smoother (EnKS)~\citep{evensen2000ensemble}, iterative EnKS~\citep{bocquet2013iterative,sakov2012iterative} and
ensemble-based four-dimensional variational method (En4D-Var)~\citep{liu2008ensemble,gustafsson2014four} have been explored.
We refer the reader to \citet{carrassi2018data} for a review of common DA methods. 

\begin{figure}[t]
    \centering
    \includegraphics[scale=1]{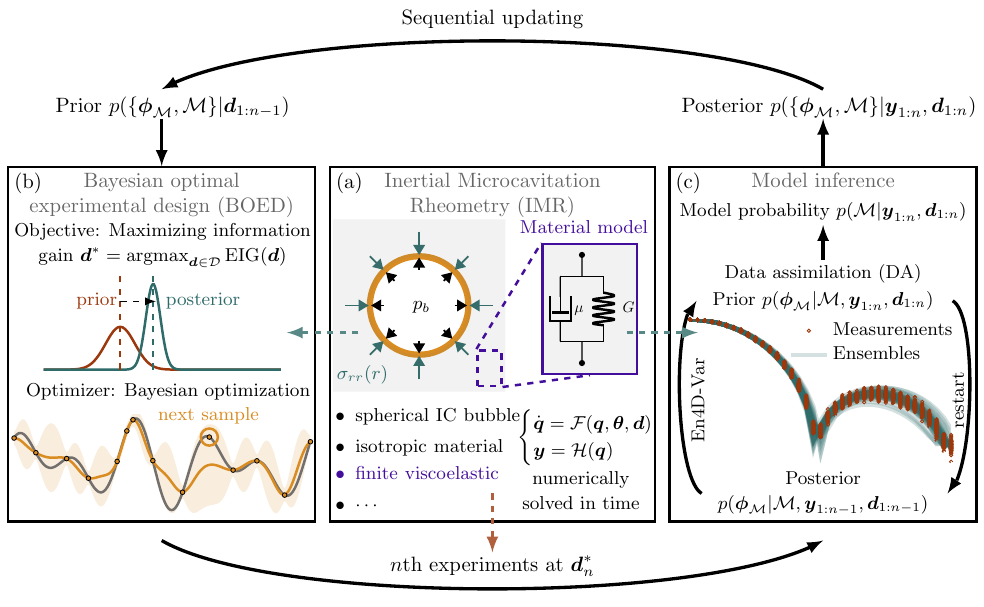}
    \caption{
        Schematic of the IMR-based sequential BOED. 
        Given a modeling parameter, $\vb*{\theta}=\{\mM,\, \vb*{\phi}_{\mM}\}$, which includes a constitutive model and its material properties, and a design $\vb*{d}$ that describes the experimental setup (for example, the equilibrium radius), the IMR approach numerically solved the spherically symmetric motion of bubble dynamics.
        In computation, the complete flow states $\vb*{q}$ include bubble radius, bubble-wall velocity, temperature, and other variables, but they are only partially observable and are denoted as $\vb*{y}$.
    }
    \label{f:overview}
\end{figure}

\citet{spratt2021characterizing} incorporated ensemble-based DA methods with the IMR solver to provide a scalable bubble-collapse rheometry framework.
It reduces the number of simulations required for accurate characterizations from a large volume in the brute-force curve fitting strategy in \citet{estrada2018high} to 48~ensembles, offering computational advantages.
The hybrid En4D-Var method also requires fewer measurements per data set to characterize the mechanical properties of hydrogels~\citep{spratt2021characterizing,mancia2021acoustic}.
DA methods require appropriate theoretical models as a \textit{prior}, yet they do not provideinformation about how to select them.
The most direct approach is to choose the model that minimizes the least-squares error. 
However, this approach does not consider the uncertainty from different sources, such as experimental setups or measurement errors. 
To address this, we use the Bayesian model selection framework~\citep{wasserman2000bayesian,chipman2001practical} to
systematically determine the model probability in the presence of these uncertainties.
We consider a library of potential constitutive models and calculate the likelihood of each model using the associated posterior distributions. 
The model parameters and probabilities are used to establish the sample space and determine the optimal EIG for optimal design.
This approach systematically and sequentially infers the model from measurements via different experimental designs.

\Cref{f:overview} shows a graphical overview of the sequential BOED procedure we use herein.
In \cref{S:IMR}, we introduce the IMR method and use it to conduct efficient bubble cavitation simulations that provide the full flow states, $\vb*{q}$.
The optimal design process in \cref{S:BOED} maximizes the EIG using BO to design the most informative cavitation experiments, denoted by $\vb*{d}^\star$.
The model inference part in \cref{S:DA} characterizes the unknown material properties, $\vb*{\phi}_{\mM}$, of each constitutive model, $\mM$, by analyzing the bubble dynamics trajectories, $\vb*{y}$, using En4D-Var.
Taken together, they form the modeling parameter, $\vb*{\theta}=\{\mM,\, \vb*{\phi}_{\mM}\}$, that describes the constitutive behavior of the soft material.
Subsequently, the marginal likelihood is used to calibrate the model probability.
When the prior is updated using the posterior, one iteration of the sequential design is completed. 
Soft material properties are shown to be accurately and efficiently characterized by iterating optimal design and model inference processes.
The performance of the sequential approach is demonstrated in \cref{S:results}
using two synthetic data sets for stiff and soft PA hydrogels.
\Cref{s:limits,s:conclusions} summarizes the main contributions and limitations.

\section{Methods}\label{S:IMR}

\subsection{Theoretical bubble dynamics model}

Different spherical bubble dynamics models have been explored in the pursuit of characterizing the viscoelastic properties of surrounding materials; cavitation in soft materials is one prominent example~\citep{gaudron2015bubble,yang2005model,estrada2018high,wilson2019comparative,barajas2017effects}. 
In these models, the Keller--Miksis equation~\citep{keller1980bubble} is applied to govern the spherically symmetric motion of bubble dynamics in a viscoelastic material assumed to be nearly incompressible. 
Upon nondimensionalization using the maximum bubble radius, $R_{\textrm{max}}$, the far-field pressure, $p_{\infty}$, the surrounding material density $\rho$, and the far-field temperature $T_{\infty}$,
the dimensionless Keller--Miksis equation is
\begin{align}
    \left(1-\frac{\dot{R}^*}{c^*}\right)R^*\Ddot{R}^*+\frac{3}{2}\left(1-\frac{\dot{R}^*}{3c^*}\right)\dot{R}^{*^2}=\left(1+\frac{\dot{R}^*}{c^*}+\frac{R^*}{c^*} \dv{}{t}\right)\left(p_b^*-\frac{1}{\mathrm{We} \, \dot{R}^*} +S^*-1\right).
\end{align}
The details of dimensionless parameters are summarized in \cref{dimensionless_quantity}. 
The bubble contents are assumed to consist of two components: water vapor and gas considered to be non-condensible, characterized by gas constants $R_v$ and $R_g$, on the time scales of inertial cavitation~\citep{akhatov2001collapse,nigmatulin1981dynamics}.
This mixture is assumed to be homobaric and follow the ideal gas law and the 
time-dependent pressure inside the bubble, $p_b^*(t)$, is coupled to the energy equation~\citep{barajas2017effects,estrada2018high}.
We assume that the mass and heat transfer of the gases within the bubble obeys Fick's law and Fourier's law.
By neglecting the initial bubble growth phase, the laser-induced cavitation model
begins when the bubble reaches its maximum radius and thermodynamic equilibrium, $R^*(0)=1$.

\begin{table}[ht!]
\centering
\caption{Dimensionless quantities used in this manuscript.}\label{dimensionless_quantity}
{\setlength{\tabcolsep}{9pt}
\begin{tabular}{r l l}
Dimensional & Dimensionless quantity  & Quantity name       \\ \midrule
     &  $U_c=\sqrt{p_{\infty}/\rho}$     & Characteristic velocity  \\
     &  $\lambda = R/R_{\infty}$ & Material stretch ratio   \\
$t$  &  $t^*=t U_c/R_\text{max}$     & Time  \\
$R$  &  $R^*=R/R_\text{max}$ & Bubble-wall radius   \\
$U$  &  $U^*=U/U_c$ & Bubble-wall velocity   \\
$R_{\infty}$  &  $R_{\infty}^*=R_{\infty}/R_\text{max}$ & Equilibrium bubble-wall radius   \\
$c$  &  $c^*=c/U_c$ & Material wave speed   \\
$p_b$  &  $p_b^*=p_b/p_{\infty}$ & Bubble-wall pressure   \\
$p_{\text{v,\,sat}}(T_{\infty})$  & 
$p_{\text{v,\,sat}}^* = p_{\text{v,\,sat}}(T_{\infty})/p_{\infty}$ & Vapor saturation pressure \\ 
$C$  &  $C^*=1/(1+( p_b^*/p^*_{\text{v,\,sat}}-1))R_v/R_g$ &  Vapor concentration   \\
$T$  &  $T^*=T/T_{\infty}$ &  Temperature   \\
$R_\text{max}$  &  $\mathrm{We}=p_{\infty}R_\text{max}/(2\gamma)$ & Weber number   \\
$S$  &  $S^*=S/p_{\infty}$ & Stress integral   \\
$G$  &  $1/\mathrm{Ca}=G/p_{\infty}$ & 1/Cauchy number   \\
$\mu$  &  $1/\mathrm{Re}=\mu/(\rho U_c R_{\text{max}})$ & 1/Reynolds number   
\end{tabular}
}
\end{table}

While the Keller--Miksis equation accurately describes spherical bubble dynamics to first order in the Mach number, appropriate constitutive relations are necessary to model the dynamic behavior of the surrounding media in terms of the time-dependent stress integral, $S^*(t)$.
Combinations of springs and dashpots, such as the Kelvin--Voigt and Maxwell models, are often used to account for the change in strain rate throughout the bubble expansion-collapse life cycles during an inertial microcavitation event.
We aim to develop a systematic method for selecting appropriate viscoelastic constitutive models for different gel specimens.
To this end, we examine a range of constitutive models for the surrounding media, as described in \cref{database_params}. 
Specifically, the Kelvin--Voigt model, incorporating either just a Neo-Hookean elastic term~\citep{gaudron2015bubble} or an additional second-order strain-stiffening term~\citep{yang2020extracting}, often better represents the nonlinear viscoelastic behavior at high strain rates~\citep{yang2022mechanical}.
These two models are different orders of Taylor expansion of the more general Fung model~\citep{fung2013biomechanics}.
More models are available, but they are beyond the scope of this work.
The stress integral associated with the quadratic law Kelvin--Voigt (qKV) model is
\begin{align}\label{stress integral}
    S^*_{} &= \overbrace{\overbrace{-\frac{4 U^*}{\mathrm{Re} R^*}-\frac{1}{2\mathrm{Ca}_{\infty}}\left[5-\frac{4}{\lambda}-\frac{1}{\lambda^4}\right]}^{\text{Neo-Hookean Kelvin--Voigt}}
    +\frac{\alpha}{\mathrm{Ca}_{\infty}}\left[\frac{177}{20}+\frac{1}{4\lambda^8}+\frac{2}{5\lambda^5}-\frac{3}{2\lambda^4} +\frac{2}{\lambda^2}-\frac{6}{\lambda} -4\lambda \right]}^{\text{quadratic law Kelvin--Voigt}},
\end{align}
where $\alpha$ represents strain stiffening when positive and strain softening when negative~\citep{knowles1977finite}.
A lower bound of $\alpha$ to maintain positive strain energy is $\alpha\geq -2/\left(2\lambda^2+1/\lambda^4-3\right)$.
When $\alpha=0$, \cref{stress integral} reduces to the same form of the stress integral for the Neo-Hookean model, in which dynamic shear moduli are used instead of the quasistatic moduli to account for the strain stiffening effect during cavitation.
For more details, readers are referred to \citet{estrada2018high}.
Recently, \citet{mancia2021acoustic} proposed a generalized variant of qKV (Gen.\ qKV) that extends the capability to accommodate variations in the ground-state shear modulus, $G_{\infty}$, traditionally considered constant in the qKV model.
Later, we will adopt this Gen.\ qKV model to account for the measurement error in the quasistatic shear modulus.

\begin{table}[ht!]
\centering
\caption{Summary of constitutive models under consideration.}\label{database_params}
\begin{tabular}{r l l}
Model $\mM$  	 & Description  & Material properties $\vb*{\phi}_{\mM}$   \\\midrule
  &  Newtonian Fluid     & $1/\mathrm{Re}$ \\
NHE~\citep{toyjanova20143d}  &  Neo-Hookean Elastic & $1/\mathrm{Ca}$\\
NHKV~\citep{gaudron2015bubble} &  Neo-Hookean Kelvin--Voigt  & $1/\mathrm{Re},\,1/\mathrm{Ca}$ \\
qKV~\citep{yang2020extracting} & Quadratic Law Kelvin--Voigt    & $ 1/\mathrm{Re},\, \alpha,\,1/\mathrm{Ca}_{\infty}$  \\
Gen.\ qKV~\citep{spratt2024numerical} & Generalized qKV    & $1/\mathrm{Re},\,\alpha,\,1/\mathrm{Ca}_{\infty}$  \\
\end{tabular}
\end{table}

We use the modeling parameter 
\begin{align}
   \vb*{\theta} &\equiv \{\mM,\, \vb*{\phi}_\mM\},
\end{align}
to represent a candidate mathematical constitutive model and its material properties.
The design parameter is
\begin{align}
     \vb*{d} & \equiv \{\text{We},\, R_{\infty}^*\},
\end{align}
representing the experimental free parameters. 
Following \citet{estrada2018high}, in the physical context of interest, we regard densities, pressures, and temperatures as constants, though this is not a restriction of the method.
We use IMR to simulate forward-time bubble dynamics with known error signatures, which we represent as Gaussian noise in the model error and measurement noise.



\subsection{Numerical methods}


The state vector is
\begin{align}\label{eq:state}
    \vb*{q}(t) = \{R^*,\,\dot{R}^*,\, p_b,\,S^*,\,T^*,\,C^*,\,1/{\mathrm{Ca}},\,1/{\mathrm{Re}},\,\alpha \},
\end{align}
where the state parameters represent the bubble-wall radius, velocity, bubble pressure, stress integral, the discretized temperature and vapor concentration fields inside the bubble, the reciprocal-Cauchy and reciprocal-Reynolds numbers, and the strain-stiffening parameter.
The discrete-time nonlinear dynamical system takes the form of 
\begin{subequations}\label{IMR_det}
\begin{align}
    \vb*{q}_{k+1} &= \mF_k(\vb*{q}_k,\vb*{d}), \tag{6a}\\
     R^*_{k+1} &= \mH(\vb*{q}_{k+1}), \tag{6b}
\end{align}
\end{subequations}
where $\mF_k$ is the nonlinear operator given the time steps, and $\mH$ is the linear observation function that maps the state $\vb*{q}$ to a point in measurement space. 
In this study, we designate the bubble radius $R^*$ as the primary observable variable due to its direct measurability in experimental setups.
For a given time interval $t\in\left[0,T \right]$ with $N_t$ time steps, the deterministic model outputs, $\btQ = \mqty[ \vb*{q}_1 & \cdots & \vb*{q}_{Nt}]$, and the corresponding bubble dynamics measurements, $\btY = \mqty[ R^* & \cdots & R^*_{Nt}]$, can be collected as
\begin{align}\label{IMR_det_all}
    \btQ = \mF(\vb*{\theta},\vb*{d}) 
    \quad \text{and} \quad
    \btY = \mH(\btQ), 
\end{align}
where $\mF$ represents the nonlinear operator that creates the space-time states at all time instances.

Following a procedure similar to \citet{freund2015quantitative}, we incorporate the deterministic IMR solver in \cref{IMR_det_all} with the model error $\vb*{\epsilon}_m$ and the experimental error $\vb*{\epsilon}_e$ to approximate experimental measurements, such that 
\begin{subequations}\label{IMR_noisy_all}
\begin{align}
    \vb*{Q}_m &= \btQ + \vb*{\epsilon}_m =\mF(\vb*{\theta},\vb*{d}) + \vb*{\epsilon}_m, \quad \text{where} \quad \vb*{\epsilon}_m\sim\mathcal{N}(\vb*{0},\vb*{\Sigma}_m), \quad \text{and} \tag{8a}\\
    \vb*{Y} &  = \vb*{Y}_m + \vb*{\epsilon}_e = \mH(\vb*{Q}_m)+ \vb*{\epsilon}_e , \quad \text{where} \quad \vb*{\epsilon}_e\sim\mathcal{N}(\vb*{0},\vb*{\Sigma}_e). \tag{8b}
\end{align}
\end{subequations}
The observation function $\mH$ is linear, so \cref{IMR_noisy_all} can be written as 
\begin{align}\label{IMR_noisy_Y}
    \vb*{Y}  = \btY + \vb*{\epsilon}= \mH\circ\mF(\vb*{\theta},\vb*{d})+ \vb*{\epsilon} , \quad \text{where} \quad \vb*{\epsilon}\sim\mathcal{N}(\vb*{0},\vb*{\Sigma}),
\end{align}
where $\vb*{\epsilon}$ is the combined error from the model and the experiments, and the true bubble dynamics, $\btY$, are unobtainable from the measurements.
In the following, \cref{IMR_noisy_Y} is used to create synthetic measurements.

\section{Simulation-based Bayesian optimal experimental design} \label{S:BOED}

The goal of the optimal design procedure is to find a design point, $\vb*{d}^\star$, within a given design space $\mathcal{D}$ that maximizes the expectation of a utility function, $u(\vb*{d},\vb*{Y},\vb*{\theta})$.
That is,
\begin{align}
    \vb*{d}^\star & 
    = \argmax_{\vb*{d}\in\mathcal{D}} \mathrm{E}\{u(\vb*{d},\vb*{Y},\vb*{\theta})\}
    = \argmax_{\vb*{d}\in\mathcal{D}} \int_{\mathcal{Y}}\int_{\Theta} u(\vb*{d},\vb*{Y},\vb*{\theta}) p(\vb*{\theta\vert \vb*{d}},\vb*{Y}) p(\vb*{Y}\vert \vb*{d}) \,\dd\vb*{\theta} \,\dd\vb*{Y},
\end{align}
where $\mathcal{Y}$ and ${\Theta}$ represent the parameter spaces for the observations and model parameters.
The inference of parameters $\vb*{\theta}$ can be obtained based on the prior distribution observations and Bayes' rule,
\begin{align}\label{Bayes}
    \underbrace{p(\vb*{\theta}\vert \vb*{d},\vb*{Y}} )_{\text{Posterior}}
 = \frac{ \overbrace{p(\vb*{Y}\vert\vb*{\theta}, \vb*{d})}^{\text{Likelihood}}\,  \overbrace{p(\vb*{\theta}\vert \vb*{d})}^\text{Prior}}{\underbrace{p(\vb*{Y}\vert\vb*{d})}_\text{Evidence}  }.
\end{align}
The probability $p(\vb*{\theta})$ can be separated as
\begin{align}
    p(\vb*{\theta}) = p(\mM)p(\vb*{\phi}_{\mM}\vert \mM),
\end{align}
which contains the probability of the mathematical constitutive model $\mM$ and the probability of the corresponding material parameters.
From \cref{IMR_noisy_Y}, the likelihood function is
\begin{align}
    p(\vb*{Y}\vert\vb*{\theta}, \vb*{d}) 
    = \frac{1}{\sqrt{(2\pi)^{N_t} |\vb*{\Sigma}|}} \mathrm{exp}\left[ -\frac{1}{2}(\vb*{Y} -\mH\circ\mF(\vb*{\theta},\vb*{d}) )\vb*{\Sigma}^{-1} (\vb*{Y} -\mH\circ\mF(\vb*{\theta},\vb*{d}) )^\top\right],
\end{align}
and the evidence is obtained through integration as
\begin{align}
    p(\vb*{Y}\vert \vb*{d}) & = \int_{\Theta} p(\vb*{Y}\vert\vb*{\theta}, \vb*{d}) p(\vb*{\theta}) \,\dd {\vb*{\theta}}.
\end{align}
The maximum information gain from the prospective experiment follows from using a relative entropy utility function, which is the same as the Kullback--Leibler (KL) divergence between the posterior and prior~\citep{lindley1956measure}, so
\begin{align}
    u(\vb*{d},\vb*{Y},\vb*{\theta}) & 
    = \infdiv{\text{posterior}}{\text{prior}} 
    =  \int_{\Theta} p(\vb*{\theta}\vert \vb*{d},\vb*{Y}) \log{ \left[\frac{p(\vb*{\theta}\vert \vb*{d},\vb*{Y})}{p(\vb*{\theta})}\right]} \,\dd \vb*{\theta} 
    = u(\vb*{d},\vb*{Y}).
\end{align}
This choice of utility function is not a function of the parameters $\vb*{\theta}$.
The expectation of the KL divergence is then
\begin{align}\label{EIG}
 \mathrm{E}\{u(\vb*{d},\vb*{Y},\vb*{\theta})\}
   &= \int_{\mathcal{Y}} \int_{\Theta} 
    p(\vb*{\theta}\vert \vb*{d},\vb*{Y}) 
    \log{ \left[\frac{p(\vb*{\theta}\vert \vb*{d},\vb*{Y})}{p(\vb*{\theta})}\right]} 
    \,\dd \vb*{\theta}\, p(\vb*{Y}\vert\vb*{d}) \, \dd \vb*{Y} \\
   &=\int_{\mathcal{Y}} \int_{\Theta}  
    \log{ \left[\frac{p( \vb*{Y}\vert \vb*{\theta}, \vb*{d})}{p(\vb*{Y}\vert\vb*{d})}\right]} 
    p(\vb*{Y}\vert \vb*{\theta}, \vb*{d}) p( \vb*{\theta}) 
    \,\dd \vb*{\theta} \, \dd \vb*{Y},
\end{align}
where the Bayes' rule in \cref{Bayes} is applied. 
This quantity is also known as the expected information gain (EIG). 
Further, $p(\vb*{\theta}\vert \vb*{d}) = p(\vb*{\theta})$, as specifying $\vb*{d}$ does not provide further information regarding $\vb*{\theta}$. 
In practice, the double integral in \cref{EIG} cannot be computed analytically and is expensive to approximate.
To address this, a double-loop Monte Carlo (DLMC) estimator, also known as the nested MC (NMC) estimator, approximates the EIG~\citep{ryan2003estimating}. 
It is
\begin{align} \label{NMC}
     \mathrm{EIG}(\vb*{d}) 
   & \approx   \mu_{\text{NMC}}(\vb*{d})
   \equiv \frac{1}{N_2} \sum_{j=1}^{N_2}  \log{ \left[\frac{p( \vb*{Y}^{(j)}\vert \vb*{\theta}^{(0,j)}, \vb*{d})}{ \frac{1}{N_1}\sum_{i=1}^{N_1} p(\vb*{Y}^{(j)} \vert \vb*{\theta}^{(i,j)}, \vb*{d})}\right]}, 
\end{align}
where $ \vb*{\theta}^{(i,j)} \overset{\text{i.i.d.}}{\sim} p(\vb*{\theta})$ and $\vb*{Y}^{(j)}  \overset{\text{i.i.d.}}{\sim} p(\vb*{Y}\vert\vb*{\theta}^{(0,j)}, \vb*{d}) $.
The samples $\theta^{(0,j)}$ are used to approximate the outer loop integral,  while $\theta^{(i=1\to N_1,j)}$ are used in the inner loop.
To obtain the dependent pair $(\vb*{\theta}^{(i,j)}, \vb*{Y}^{(i)})$, the importance sampling technique is used: we first draw $\vb*{\theta}^{(i,j)}$ from the prior $p(\vb*{\theta})$, and then draw $\vb*{Y}^{(i)}$ from the conditional distribution $p( \vb*{Y}\vert \vb*{\theta}^{(i,j)}, \vb*{d})$.
In the computation, the samples $\vb*{\theta}^{(i,j)}$ are collected using the sample reused technique~\citep{huan2013simulation}. 
This technique uses a batch of prior samples $\{\vb*{\theta}^{(l)}\}_{l=1}^{N_2}$ for both the inner and outer Monte Carlo sums, reducing the computational cost from $O(N_1 N_2)$ to $O(N_2)$. 
In the following, we use the notation $N_\text{EIG}$ to represent the sample size used for approximating the EIG.

In practical settings, experiments and data collection for inertial cavitation are carried out separately due to the need to prepare hydrogel specimens for different experimental setups. 
Thus, a sequential experimental design is important for this purpose.
We assume that the experiment outcomes are conditionally independent, given the latent variables and designs,
\begin{align}
     p(\vb*{Y}_{1:N_\text{Des}}, \vb*{\theta}\vert  \vb*{d}_{1:N_\text{Des}}) 
     = p(\vb*{\theta}) \prod_{n = 1}^{N_\text{Des}} p(\vb*{Y}_n\vert \vb*{\theta}, \vb*{d}_n).
\end{align}
Having conducted experiments $1,2,\cdots, N_\text{Des}-1$, the design $\vb*{d}_{N_\text{Des}}$ for the prospective experiment can be obtained by replacing the prior, $p(\vb*{\theta})$, with $p(\vb*{\theta}\vert \vb*{d}_{1:N_\text{Des}-1},\vb*{Y}_{1:N_\text{Des}-1})$ in \cref{Bayes} such that 
\begin{align}
    p(\vb*{\theta}\vert \vb*{Y}_{1:N_\text{Des}}, \vb*{d}_{1:N_\text{Des}}) 
    =  \frac{ p(\vb*{Y}_{N_\text{Des}}\vert\vb*{\theta}, \vb*{d}_{N_\text{Des}})\, p(\vb*{\theta}\vert \vb*{Y}_{1:N_\text{Des}-1}, \vb*{d}_{1:N_\text{Des}-1})}{p(\vb*{Y}_{N_\text{Des}}\vert\vb*{d}_{N_\text{Des}})}
    = ... 
    =  \frac{p(\vb*{\theta}) \prod_{n = 1}^{N_\text{Des}} p(\vb*{Y}_{n}\vert\vb*{\theta}, \vb*{d}_{n}) }{p(\vb*{Y}_{1:N_\text{Des}}\vert\vb*{d}_{1:N_\text{Des}})}.
\end{align}
Similar to \cref{NMC}, the EIG for $N_\text{Des}$ is approximated in a Markovian fashion as 
\begin{align}\label{eqn:EIG_sequential}
     \mathrm{EIG}(\vb*{d}_{N_\text{Des}}) 
   & \approx   \frac{1}{N_\mathrm{EIG}} \sum_{j=1}^{N_\mathrm{EIG}}  \log{ \left[
   \frac{p( \vb*{Y}_{N_\text{Des}}^{(j)}\vert \vb*{\theta}_{N_\text{Des}}^{(0,j)}, \vb*{d}_{N_\text{Des}})}
   { \frac{1}{N_\mathrm{EIG}}\sum_{i=1}^{N_\mathrm{EIG}} 
   p(\vb*{Y}_{N_\text{Des}}^{(j)} \vert \vb*{\theta}_{N_\text{Des}}^{(i,j)}, \vb*{d}_{N_\text{Des}})}\right]}, 
\end{align}
where $\vb*{\theta}_{N_\text{Des}}^{(i,j)} \overset{\text{i.i.d.}}{\sim} {p(\vb*{\theta}\vert \vb*{Y}_{1:N_\text{Des}-1 }, \vb*{d}_{1:{N_\text{Des}}-1 })}$ and $\vb*{Y}_{N_\text{Des}}^{(j)}  \overset{\text{i.i.d.}}{\sim} p(\vb*{Y}\vert\vb*{\theta}_{N_\text{Des}}^{(0,j)}, \vb*{d}_{N_\text{Des}}).$
Through this procedure, we conduct an adaptive sequential experiment that iteratively optimizes the selection of the design $\vb*{d}_{N_\text{Des}}$ at each step. 
For each such step, we solve a sequential optimization problem
\begin{align}\label{opt_seq}
    \vb*{d}_{N_\text{Des}}^\star
    = \argmax_{\vb*{d}_{N_\text{Des}}\in\mathcal{D}} \mathrm{EIG}(\vb*{d}_{N_\text{Des}}),
\end{align}
Given an EIG estimator, differnt methods can be used for \cref{opt_seq}, including some specifically developed for BOED~\citep{amzal2006bayesian, huan2013simulation, muller2005simulation}.  
Here, Bayesian optimization (BO) is selected for the subsequent design optimization, given its advantageous features such as sample efficiency, robustness to multi-modality, and inherent capability to handle noisy objective evaluations~\citep{jones1998efficient}. 
Following \citet{snoek2012practical}, we use the Ard Mat{\'e}rn $5/2$ kernel for Gaussian process (GP) regression and the expected improvement criterion for the acquisition function.
Further details are provided in \cref{App:BO}.
In practice, we initialize BO by evaluating the EIG values at $N_{\text{Int}}$ random designs.
This strategy creates a more reasonable initial GP model~\citep{kandasamy2020tuning, Benjamin2019,eriksson2021scalable}.
A total number of $N_{\text{BO}}$ BO trials is used to obtain the optimal design.

\begin{algorithm}
    \caption{
        Bayesian optimal experimental design (refer to \cref{f:overview}~(b) for graphical illustration) 
    }\label{alg:BOED}
    \hspace*{\algorithmicindent} \textbf{Input:} prior $p(\vb*{\theta}\vert \vb*{d}_{1:N_\text{Des}})$, error variance $\vb*{\Sigma}$, EIG sample size $N_\mathrm{EIG}$ \\
    \hspace*{\algorithmicindent} \textbf{Output:} Next design point $\vb*{d}_{N_\text{Des}+1}^\star$
    \begin{algorithmic}[1]
        \State Evaluate EIG for the $N_{\text{Int}}$ random points
        \For{$l=N_{\text{Int}}+1:N_\text{BO}$}
           \State Perform Gaussian process regression based on the evaluated values, $\{\mathrm{EIG}(\vb*{d}_{N_\text{Des}+1}^{(l')})\}_{l'=1}^{l}$
            \State Obtain next search point, $\vb*{d}_{N_\text{Des}+1}^{(l+1)}$, that maximizes expected improvement
            \State Evaluate $\mathrm{EIG}(\vb*{d}_{N_\text{Des}+1}^{(l+1)})$
        \EndFor
        \State  $\vb*{d}_{N_\text{Des}+1}^\star\leftarrow \argmax_{1\leq l\leq N_{\text{BO}}+1}{\{\mathrm{EIG}(\vb*{d}_{N_\text{Des}+1}^{(l)}) \}}$.
    \end{algorithmic} 
    \begin{algorithmic}[1]
        \Function{EIG}{$\vb*{d};p(\vb*{\theta}\vert \vb*{d}_{1:N_\text{Des} });N_\mathrm{EIG}$} 
            \State Draw $N_\mathrm{EIG}+1$ samples $\left(\vb*{\theta}^{(0)},\vb*{\theta}^{(1)},\cdots, \vb*{\theta}^{(N_\mathrm{EIG})}\right)$ from prior $p(\vb*{\theta}\vert \vb*{d}_{1:N_\text{Des} })$ 
                \State Draw $N_\mathrm{EIG}$ samples $\left(\vb*{Y}^{(1)},\cdots, \vb*{Y}^{(N_\mathrm{EIG})}\right)$ from the likelihood $p(\vb*{Y}\vert \vb*{\theta}^{(0)}, \vb*{d})$ with Gaussian error $\vb*{\Sigma}$
            \For {$i = 1:N_\mathrm{EIG}$} 
            \State Perform the IMR simulation for the design $\vb*{d}$ using the parameter $\vb*{\theta}^{(i)}$
            \State Evaluate the likelihood $p(\vb*{Y}^{(j)}\vert \vb*{\theta}^{(i)}, \vb*{d})$ with Gaussian error $\vb*{\Sigma}$
            \EndFor
            \State Calculate the EIG using \cref{eqn:EIG_sequential}
        \EndFunction
    \end{algorithmic}
\end{algorithm}

An algorithm for IMR-based BOED is outlined in \cref{alg:BOED}. 
This strategy is systematic and identifies the optimal design for the next experiment. 
The next step involves characterizing the material properties based on the measurements of bubble dynamics.

\section{Model inference}\label{S:DA}

\subsection{Data assimilation}

With data collected from experiments or simulations on inertial cavitation, the remaining task is to find the most accurate model for characterizing bubble dynamics within uncertainty-prone high-strain-rate regimes.
Here, we adopt the En4D-Var approach due to its computational efficiency~\citep{spratt2021characterizing,mancia2021acoustic}.
We assume the variables follow a multivariate Gaussian distribution and use $N_\text{En}$ ensembles, $\btQ_0 =\left( \btQ_0^{(1)}, \cdots, \btQ_{0}^{(N_{\text{En}})}\right)$, to approximate this distribution based on a given observed data set, $\vb*{Y}^{\mathrm{D}}$, and a data assimilation window size.
Details of the standard En4D-Var method are provided in \cref{App:En4dVar}, along with three enhancements introduced here.
First, the reciprocal-Cauchy and reciprocal-Reynolds numbers, $1/\mathrm{Ca}$ and $1/\mathrm{Re}$, are incorporated into the state vector in \cref{eq:state} to guarantee Gaussian distributions of the physical quantities, $G$ and $\mu$. 
Second, the parameter $\alpha$ can be negative, corresponding to strain-softening, when the quasistatic shear modulus, $G_{\infty}$, is overestimated. 
Third, instead of performing En4D-Var for every measurement independently and then collecting all the posterior ensembles, we consider an iterative-restart strategy to reduce the computational cost and bias from the prior.
A similar restart strategy has been used in the restart-EnKF to address the dynamical systems with strong nonlinearity~\citep{zafari2005assessing, gu2007iterative, hendricks2008real}.
We apply En4D-var to the data mean, and the measurement noise matrix $\vb*{P}_k$ at each time step is obtained from the data.
After obtaining the posterior ensembles, we restart the data assimilation process by drawing fresh samples from the inflated posterior distribution. 
Here, the ``Relaxation Prior to Spread''~(RTPS) scheme addresses the sampling error in ensemble methods due to finite ensemble size~\citep{whitaker2012evaluating}. 
The variances are updated as 
\begin{align}
    \sigma_i = 
        {\sigma_i^{\text{(post)}}} + a \left(
            {\sigma_i^{\text{(prior)}}-\sigma_i^{\text{(post)}}}
        \right), 
\end{align}
where $a = 0.7$ is an inflation parameter~\citep{spratt2021characterizing}. 
We repeat this process, and the final posterior distributions are obtained through $N_\text{runs}$ complete cycles. 
Thus, the total number of DA runs required is $N_\text{DA} = N_{\text{iter}} N_\text{runs}$.

\subsection{Model probability}

After performing data assimilation for available models, the next step is to choose models that best represent the experimental measurements.
The most straightforward way is to select the model with the least-squares error.
This strategy, however, does not account for the uncertainty in measurements.
To tackle this, we calculate the probability of each model from the En4D-Var outputs using the Bayesian model selection framework~\citep{wasserman2000bayesian,chipman2001practical}.
Given the measurement data $\vb*{Y}^{\mathrm{D}}$,
the marginal likelihood of each model $\mM$ can be calculated as
\begin{align}
p(\mM \vert \vb*{Y}^{\mathrm{D}}, \vb*{d}) 
= \frac{p(\mM )}{p(  \vb*{Y}^{\mathrm{D}}\vert  \vb*{d})}
   \int_{{\Theta}} p(\vb*{Y}^{\mathrm{D}}\vert \mM, \vb*{\phi}_{\mM}, \vb*{d}) p(\vb*{\phi}_{\mM} \vert \mM, \vb*{d}) \, \dd \vb*{\phi}_{\mM} .
\end{align}
Similar to \cref{NMC}, importance sampling can be used to approximate the marginal likelihood as 
\begin{align}\label{marginal_approx}
p(\mM \vert \vb*{Y}^{\mathrm{D}} , \vb*{d}) 
\approx \frac{1}{N_\text{En}} 
\sum_{i = 1}^{N_\text{En}} p(\vb*{Y}^{\mathrm{D}}\vert \mM, \vb*{\phi}^{(i)}_{\mM}, \vb*{d}),
\end{align}
where $\vb*{\phi}^{(i)}_{\mM}\sim  p(\vb*{\phi}_{\mM} \vert \mM, \vb*{d})$.
If one assumes the models can fully represent the experiments, then $\sum_{\mM} p(\mM) =1$. 
The posterior probability of the model $\mM$ can then be normalized as 
\begin{align}
      p(\mM \vert \vb*{Y}^{\mathrm{D}} , \vb*{d})\propto  {p(\mM \vert \vb*{Y}^{\mathrm{D}} , \vb*{d})}/{\sum_{\mM}p(\mM \vert \vb*{Y}^{\mathrm{D}} , \vb*{d})}.
\end{align}
The obtained posterior distribution, 
\begin{align}
     p(\vb*{\theta}\vert\vb*{Y}_{1:N_\text{Des}},\vb*{d}_{1:N_\text{Des}} ) 
     = p(\mM\vert\vb*{Y}_{1:N_\text{Des}},\vb*{d}_{1:N_\text{Des}})
     p(\vb*{\phi}_{\mM}\vert \mM,\vb*{Y}_{1:N_\text{Des}},\vb*{d}_{1:N_\text{Des}}),
\end{align}
is subsequently used to update the prior for \cref{alg:BOED} to obtain the next optimal design point.

\begin{algorithm}
    \caption{
        Model inference (refer to \cref{f:overview}(c) for graphical illustration)
    }\label{alg:DA}
    \hspace*{\algorithmicindent} \textbf{Input:} target design $\vb*{d}$, prior distribution $p(\left(\vb*{\phi}_\mM , \mM\right) \vert \vb*{d}_{1:N_\text{Des} })$ \\
    \hspace*{\algorithmicindent} \textbf{Output:} posterior distribution  $p(\left(\vb*{\phi}_{\mM} , \mM\right) \vert \vb*{Y}^{\mathrm{D}} , \vb*{d})$
    \begin{algorithmic}[1]
        \State Collect data $\vb*{Y}^{\mathrm{D}}$ at the design $\vb*{d}$ with error $\vb*{\Sigma}$
        \For {each  model $\mM$}
        \For{$l=1:N_r$}
        \State Draw $N_{\text{En}}$ samples $\left( \vb*{\tilde{\theta}}_0^{(1)}, \cdots, \vb*{\tilde{\theta}}_{0}^{(N_{\text{En}})}\right)$ from the prior distribution \textbf{$p(\vb*{\phi}_{\mM}\vert \mM, \vb*{d}_{1:N_\mathrm{Des} })$}
           \State Generate $N_{\text{En}}$ initial ensembles $\btQ_0 =\left( \btQ_0^{(1)}, \cdots, \btQ_{0}^{(N_{\text{En}})}\right)$
           \State Perform En4D-Var with $N_\text{iter}$ iterations to update the ensembles $\btQ_0$
            \State Perform covariance inflation and update the prior distribution 
        \EndFor
        \State Calculate the marginal likelihood $p(\mM \vert \vb*{Y}^{\mathrm{D}} , \vb*{d})$ using \cref{marginal_approx}
        \EndFor
        \State Normalize the model probability to obtain the posterior distribution $p(\left(\vb*{\phi}_\mM , \mM\right) \vert \vb*{Y}^{\mathrm{D}} , \vb*{d})$
        \State Update the prior $p(\left(\vb*{\phi}_{\mM} , \mM\right)\vert \vb*{d}_{1:N_\text{Des}+1})$ for \cref{alg:BOED} to obtain the next design point
    \end{algorithmic}
\end{algorithm}
An algorithm for IMR-based model inference is presented in \cref{alg:DA}. 
Together with \cref{alg:BOED}, these form a complete loop for the simulation-based characterization of soft matter, as illustrated in \cref{f:overview}.

\section{Results} \label{S:results}


\begin{table}[ht!]
    \centering
    \caption{Summary of synthetic datasets. 
    The characterization of these parameters from experimental data are demonstrated in \citet{yang2020extracting}.}\label{datasets}
    \begin{tabular}{c c c c c c c}
     & \multirow{2}{*}{Material} &  \multirow{2}{*}{Model $\mM$} 
     &  \multicolumn{4}{c}{Parameters $\vb*{\phi}_{\mM}$} \\
    & & & $G_{\infty}$ [\unit{\kilo\pascal}] 	 & $G$ [\unit{\kilo\pascal}]  & $\mu$ [\unit{\pascal\,\second}] & $\alpha$   \\\midrule
    Case 1& Stiff PA& qKV & 2.77 &  ---    & 0.186 & 0.48  \\
    Case 2& Soft PA & NHKV & 0.57 & 8.31 &0.093     & ---  
    \end{tabular}
\end{table}

We demonstrate the proposed framework by using the IMR method to create two datasets.
The underlying models for these datasets are chosen to mimic the viscoelastic behavior of stiff and soft PA hydrogels~\citep{yang2020extracting}. 
The details are summarized in \cref{datasets}.
To align these simulation-based datasets with real-world experimental measurements, we introduce synthetic error to accommodate different sources of error. These include uncertainties in measurement errors and aliasing in the bubble response.
The standard deviations of this synthetic noise, $\sigma = |R^*-1|/50+t^*/160$, are tailored to depend on time and state, qualitatively reflecting experimental measurements~\citep{estrada2018high,yang2020extracting,yang2022mechanical}.
A longer duration of measurement or being closer to bubble collapse will result in a larger error, as illustrated in \cref{f:model selection}.
A set of measurements containing 100 $R(t)$ curves is collected for each design.
We aim to accurately characterize the underlying model with a minimum requirement of design iterations using the optimal sequential design process, as shown in \cref{f:overview}.
We consider two candidate models, NHKV and Gen.\ qKV, to demonstrate the proposed framework at a reasonable computational cost. 
The design is initialized with a probability of 50$\%$--50$\%$ for these two models.
To better represent real-world experiments, the optimization problems for the design parameters are restricted within the ranges $\mathrm{We}\in [100,1000]$ and $R^*_{\infty}\in[0.14,0.3]$~\citep{estrada2018high,yang2020extracting,yang2022mechanical,mancia2021acoustic}.
In the computation, the data assimilation window is set up to the first two peaks of the bubble collapse.
For each set of measurements, En4D-Var is run $N_{\text{runs}}=3$~times (with 2~restarts), using 5~iterations for each run and an ensemble size of $N_\text{En}=48$.
This choice of ensemble size follows \citet{spratt2021characterizing}.
Later, we will show that the above setup is enough to characterize the underlying model of synthetic data.

Computations are performed on PSC~Bridges2 using a dual AMD 64-core CPUs (SKU~7742, Rome). 
The default wall-clock time is 30 minutes, and the memory per core is \SI{1}{\giga\byte}.
The CPU hours required vary between the two models due to differences in automatic time step requirements needed to address stiffness near bubble collapse. 
Generally, a single design consisting of $N_\text{BO}=15$ BO trials with $N_{\text{EIG}}=1000$ EIG samples and its subsequent DA process requires approximately 200~CPU~hours when all simulations are performed using the qKV model, and an additional 200~CPU~hours when using the NHKV model.
Quantitative assessment metrics include EIG, root-mean-square error (RMSE) of the $R(t)$ data, and the relative error of the material properties.


\subsection{qKV for stiff PA}\label{S:stiff PA}

We first consider qKV as the underlying model to approximate the behavior of stiff PA~\citep{yang2020extracting}.
We assume that the quasistatic shear modulus can be measured with $G_{\infty}=\SI{2.77\pm0.3}{\kilo\pascal}$, which has a higher variance than the experimental measurements. 
The prior distributions of the material properties are set as $G=\SI{15.09\pm4.35}{\kilo\pascal}$, $\mu= \SI{0.209\pm0.18}{\pascal\second}$ for NHKV and $\mu= \SI{0.286\pm0.186}{\pascal\second}$, $\alpha= 0.28\pm0.48$ for Gen.\ qKV.
The latter has around $50\%$ error compared to the underlying truth, which has been shown as a reasonable offset to validate the performance of DA~\citep{spratt2021characterizing}.
Truncated Gaussian distributions \citep{robert1995simulation} ensures $\mu>0$ such that the material properties are physically interpretable. 

\begin{figure}
    \centering
    \includegraphics[scale=1]{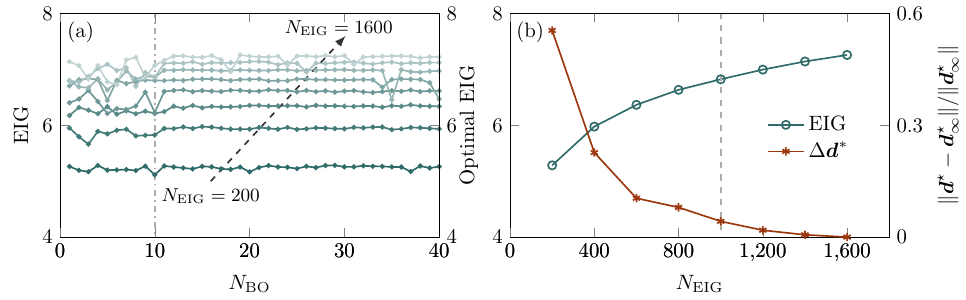}
    \caption{
        BO output trajectories (a) and the optimal BO outputs (b) for EIG sample sizes $N_\mathrm{EIG}= 200, 400, \dots, 1600$.
        The dot-dashed line in (a) indicates the onset of the BO process with 10 initial trials. 
        The relative difference between the optimal design, $\vb*{d}^*$, using $N_\mathrm{EIG}$ samples and the design using 1600 samples, $\vb*{d}_{\infty}^\star$, is shown in (b).
        The optimal EIG for the EIG sample size used later, $N_\mathrm{EIG} = 1000$, is highlighted in (b).
    }
    \label{f:BO_converge}
\end{figure}

We first show the results of the simulation-based BOED in \cref{S:BOED} using the aforementioned prior distributions as an example. 
\Cref{f:BO_converge}~(a) shows the BO outputs for different sample sizes used to estimate the EIG.
In general, the observed EIG increases as the sample size grows. 
Even without additional noise, the meaningful uncertainty in the initial model selection and material properties leads to a potential for gaining information through experiments. 
As a result, the EIG values are comparable for the same EIG sample size.
With an initialization of 10 random trials, only a few more trials are necessary to reach the optimal EIG values. 
These values are shown in \cref{f:BO_converge}~(b) for different sample sizes, where a decreasing trend in the slope can be observed.
A similar trend can be observed for the relative error in the optimal design parameters.
Note that the EIG serves as a guiding variable for identifying the optimal design, and its actual value is not meaningful in this context.
Based on these observations, we will estimate the EIG using a sample size of $N_\mathrm{EIG}=1000$ and perform $N_\mathrm{BO}=15$ trials for BO in the design process to achieve a reasonable balance between accuracy and computational efficiency.

\begin{figure}[ht]
    \centering
    \includegraphics[scale=1]{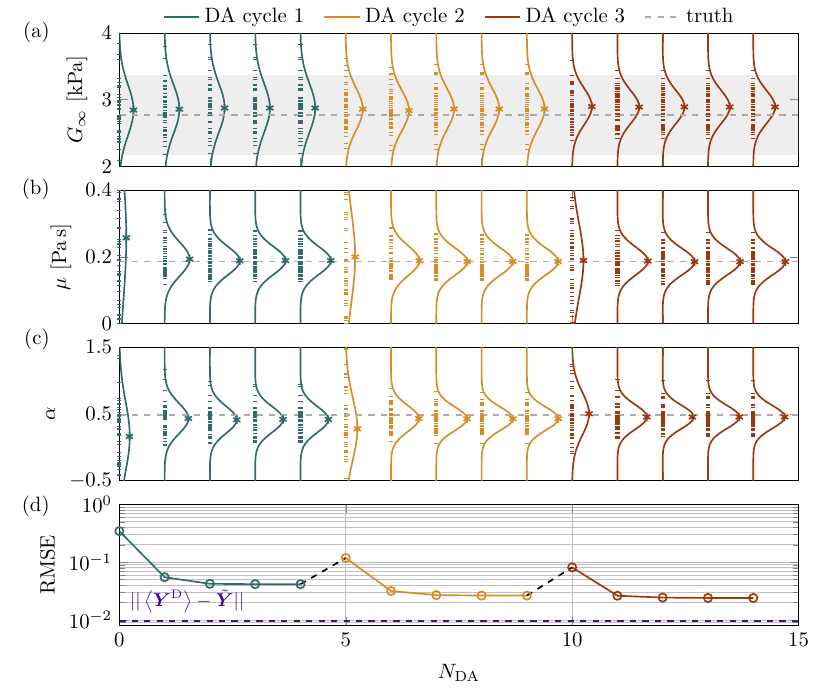}
    \caption{
        DA outputs over the total En4D-var iteration number $N_\text{DA}$: ensembles for (a) $G_{\infty}$; (b) $\mu$; (c) $\alpha$; and (d) RMSE of the bubble dynamics curves (see examples in~\cref{f:model selection}). 
        The shaded area in (a) represents the $95\%$ confidence interval for the $G_{\infty}$ measurements.
        The solid curves in~(a--c) represent Gaussian distributions approximated from the 48 ensembles, with their respective mean values marked as stars.
        In~(d), the error between the mean of the measurements and the unobtainable truth, $\|\langle\vb*{Y}^{\mathrm{D}}\rangle-\btY\|$, is shown for comparison.
    }
    \label{f:DA_params_case1}
\end{figure}

Next, we collect measurements at the optimal design and perform data assimilation to obtain the posterior distributions. 
For example, \cref{f:DA_params_case1} shows the DA outputs using the initial prior distributions for Gen.\ qKV.
As expected, the variance of the ensembles decreases with more DA iterations.
Despite an initial guess of approximately 50$\%$ error, using En4D-Var enables accurate identification of the true material properties. 
It can be observed that the restart strategy with covariance inflation enhances the posterior distributions with more restart runs, leading to a decrease in the RMSE of the ensemble $R(t)$ curves.
Compared to the standard En4D-Var in \citet{spratt2021characterizing}, drawing fresh samples when restarting helps avoid local minima due to initial bias in ensembles.

\begin{figure}
    \centering
    \includegraphics[scale=1]{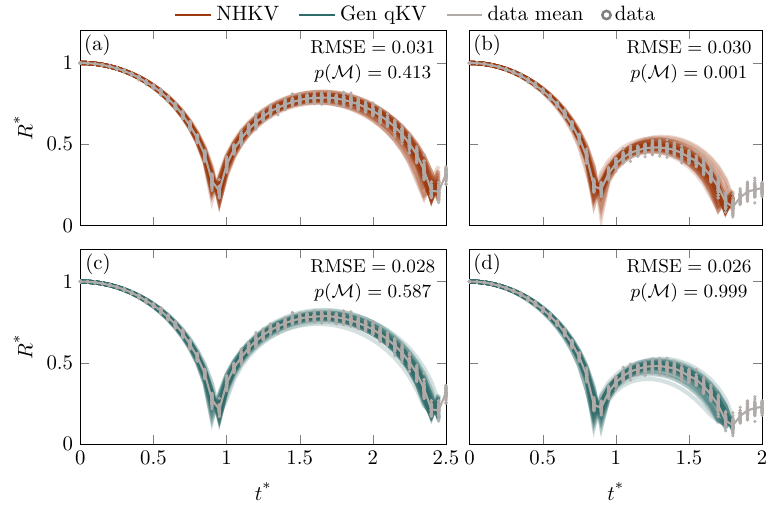}
    \caption{
    Posterior bubble dynamics trajectories and their marginal likelihoods: (a, c) $R_\text{max}=\SI{9.85d-4}{\meter}$ and $R_{\infty}^*=0.2887$; (b, d) $R_\text{max}=\SI{3.87d-4}{\meter}$ and $R_{\infty}^*=0.15$.
    }
    \label{f:model selection}
\end{figure}

The final step of the sequential design is to calibrate the model probabilities based on the measurements and posteriors.
\Cref{f:model selection} shows the RMSE and the model probabilities by comparing the posterior $R(t)$ curves to the measurements for two designs. 
For each design, both models show favorable bubble dynamics compared to the average of measurements, resulting in similar RMSE. 
Still, the likelihoods of these two models offer a different perspective for model selection by considering the variance present in these data.
For the design in \cref{f:model selection}~(a,c), NHKV and Gen.\ qKV show comparable model probability. 
Conversely, for the other case in \cref{f:model selection}~(b, d), the preference for Gen.\ qKV over NHKV is unequivocal. 
These findings are also visually corroborated.
Magnified regions near the second bubble collapse, where the differences between the two models are most pronounced, are provided in~\cref{f:model selection_app}.
These model probabilities are next used to update the prior distribution to estimate the optimal EIG for the next design point. 
The processes shown in \crefrange{f:BO_converge}{f:model selection} are repeated.

\begin{figure}[ht]
    \centering
    \includegraphics[scale=1]{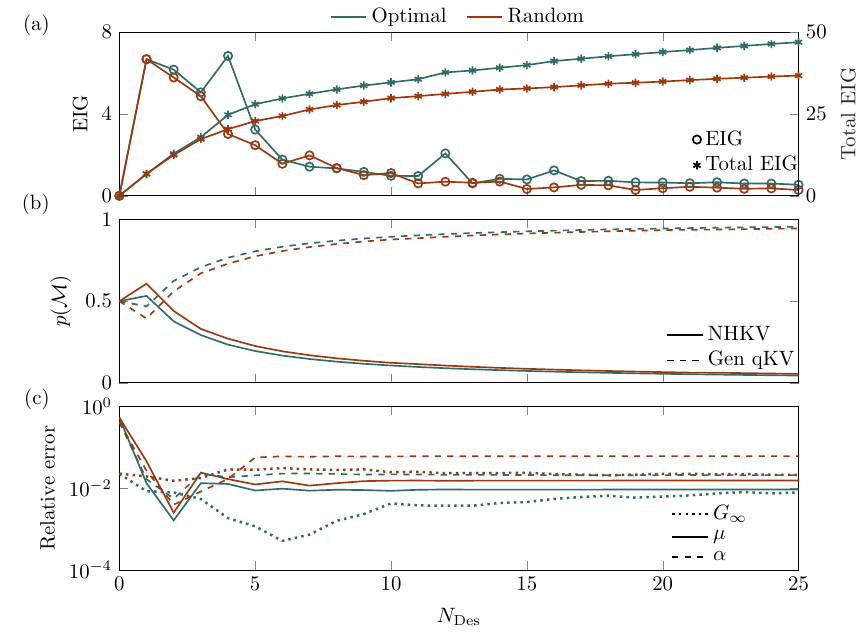}
    \caption{
        Sequential BOED outputs over the design number $N_\mathrm{Des}$:
        (a) EIG and total EIG; (b) model probabilities; and (c) relative error of the mean material properties. 
    }
    \label{f:params_case1}
\end{figure}

\Cref{f:params_case1} shows the results for the sequential BOED. 
As the number of measurements increases, we observe a trend of increased exploration of parameter space and higher model probability for Gen.\ qKV, leading to a decreasing EIG.
Conversely, the total EIG continues to rise due to inherent measurement uncertainties. 
The initial EIG values for both the optimal and randomly selected design parameters, which follow a uniform distribution, are notably high, reflecting a meaningfully large discrepancy between our chosen prior distribution and the actual underlying distribution.
As a result, experiments on any design yield substantial knowledge gains, leading to a larger EIG.
Accurate identification of the material properties for Gen.\ qKV can be seen from \cref{f:params_case1}~(c) in terms of the relative error.
Here, the distributions of material properties are cumulatively updated to incorporate the results from all the DA analyses up to the $N_{\text{Des}}$th simulation.
The mean material properties converge across approximately 10 designs, coinciding with a reduction in their variances to levels deemed negligible (see the posterior entropy shown in~\cref{f:sequential_results}~(b)).
The convergence of $G_{\infty}$ to the ground truth suggests that the Gen.\ qKV model effectively reduces to the standard qKV model with a constant quasistatic shear modulus.
These findings indicate that the sequential approach effectively characterizes the underlying qKV model despite multiple sources of error.
Compared to the random design, the optimal sequential BOED demonstrates superior EIG, model probability, and relative error performance, showing that the proposed approach can accurately and efficiently characterize the underlying soft material from bubble dynamics.

\begin{figure}[h]
    \centering
    \includegraphics[scale=1]{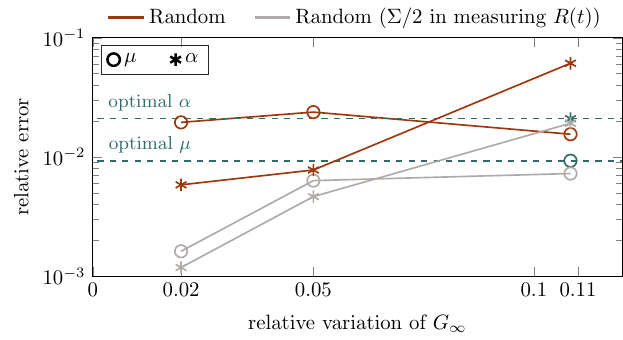}
    \caption{
        Convergence study of different error sources.
    }
    \label{f:convergence_qKV}
\end{figure}

Finally, we examine the effects of different error sources within the system, as shown in \cref{f:convergence_qKV}.
The accuracy of the stiff-straining parameter, $\alpha$, improves as the variation of the quasistatic shear modulus, $G_{\infty}$, decreases to the real experimental error of $2\%$, see, e.g., \citet{estrada2018high}.
At the same time, these parameters collectively represent a material's resistance to shearing deformation under shearing stress (see \cref{stress integral}).

A more accurate determination of the viscosity, $\mu$, requires smaller measurement errors in the bubble radius, $R$. 
This correspondence can also be inferred from \cref{stress integral} due to the coupled contributions of $\mu$ and $R$ to the stress integral.
For example, the optimal design is conducted considering high measurement noise in both $R$ and $G_{\infty}$, and the outputs demonstrate notable improvement compared to the random design.
By reducing the error in both sources, we anticipate accurately identifying the two parameters with a relative error of approximately $0.1\%$, as is the case for the random design.

\subsection{NHKV for soft PA}\label{S:soft PA}

Next, we consider NHKV as the underlying model to approximate the behavior of soft PA~\citep{yang2020extracting}.
Consistent with the previous case, we initialize the prior distributions of the NHKV material properties as $G= \SI{12\pm 6.35}{\kilo\pascal}$ and $\mu= \SI{0.14\pm0.073}{\pascal\second}$, resulting in a $50\%$ error against the truth.
For Gen.\ qKV, we assume that the quasistatic shear modulus is measured with $G_{\infty}= \SI{0.57\pm 0.06}{\kilo\pascal}$ and the prior distributions are set as $\mu= \SI{0.08\pm0.05}{\pascal\second} $, $\alpha= 0.96\pm0.48$. 

\begin{figure}[h]
    \centering
    \includegraphics[scale=1]{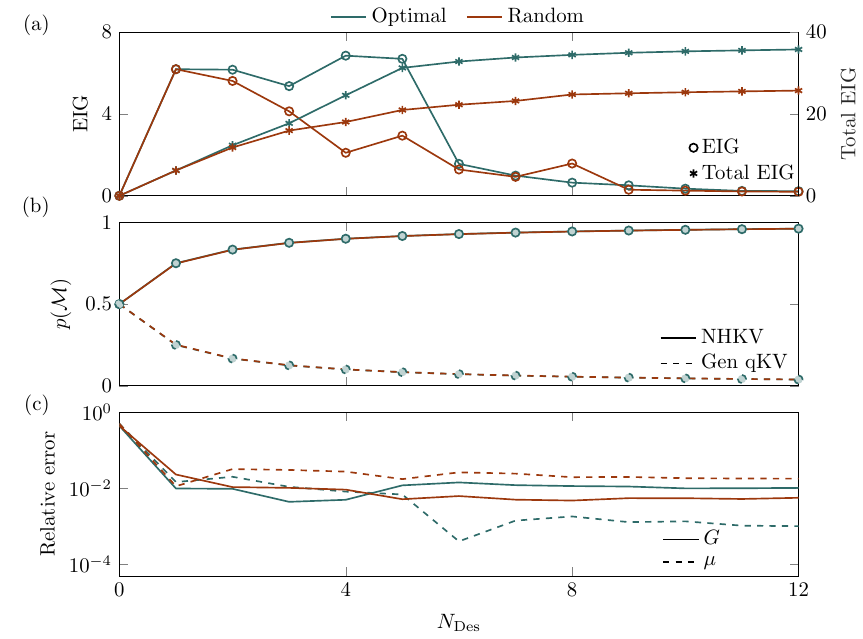}
    \caption{
        Sequential BOED outputs over the design number $N_\mathrm{Des}$: (a) EIG and total EIG; (b) model probabilities $p(\mM)$; and (c) relative error of the mean material properties.
    }
    \label{f:params_case2}
\end{figure}

We repeat the process shown in \cref{S:stiff PA} and show the sequential BOED results for the synthetic soft PA in \cref{f:params_case2}.
The overall trend is qualitatively similar to those presented in \cref{f:params_case1}.
These 12 designs accurately characterize the underlying NHKV model and its material properties.
Although the optimal designs are chosen by maximizing information gains instead of minimizing errors in material properties, they yield improved results for $\mu$ and comparable outcomes for $G$ relative to the random design.
Collectively, \cref{S:stiff PA,S:soft PA} illustrate that the proposed method can accurately and efficiently characterize the mechanical behaviors of different soft materials.

\section{Limitations of present work}\label{s:limits}

The application of En4D-Var for data assimilation achieves computational efficiency if the material properties, such as shear modulus and viscosities, follow a multivariate normal distribution.
Consequently, its performance deteriorates when the soft materials under characterization do not adhere to this assumption.
Other Bayesian parameter inference methods, such as Markov chain Monte Carlo (MCMC) sampling, can address this issue but often require many samples to compute posterior estimates with acceptable accuracy.
As suggested by \citet{kruschke2014doing}, a minimum sample size for an effective MCMC process is $10^4$, higher than the En4D-Var ensemble size used in this work, $N_\text{En}=48$.
Therefore, balancing the number of measurements required for posterior sampling and the constraints imposed on the distributions of the material properties becomes necessary for analyzing real experimental data.
Conducting a prior assessment of the test samples could potentially aid in achieving this balance.

The proposed approach necessitates knowledge of the underlying theoretical models as a \emph{prior} for optimal design and parameter inference.
Specifically, in our context, this information includes the constitutive models used within the spherical bubble dynamics equations.
While the modal probability calculation yields the marginal likelihoods for each constitutive model under consideration, it does not provide further insights beyond these models.
If all the available models inadequately represent the experimental measurements, data-driven modeling approaches, such as system identification or operator inference methods, offer a viable strategy for exploring alternative models.

\section{Conclusions}\label{s:conclusions}

This study presents a computational approach for the optimal design of experiments to accelerate the discovery of material properties.
To create synthetic data that aligns with real experiments, we used inertial microcavitation rheometry (IMR) for accurate and efficient bubble dynamics simulations.
By incorporating appropriate noise to account for model error and measurement noises, these simulations serve as predictions of bubble dynamics trajectories under specific experimental conditions during the optimal design phase and as synthetic measurements during the parameter inference phase.
We formulated the optimization problem within a Bayesian statistical framework
to design experiments that provide the most informative data about unknown material properties.
The constitutive models and associated viscoelastic properties are then determined from the measurements using a hybrid ensemble-based 4D-Var method (En4D-Var).
By iterating these two processes sequentially, we demonstrated accurate and efficient characterizations of two types of synthetic polyacrylamide (PA) gels.
The larger error in each source of synthetic data compared to real experimental measurements evidences the robustness of the IMR-based design approach, underscoring its potential applicability to actual experimental designs.

\section*{CRediT authorship contribution statement}

\textbf{TC}: Formal analysis, Methodology, Software, Investigation, Data Curation, Validation, Visualization, Writing – original draft, Writing – review $\&$
editing.
\textbf{JBE}: Conceptualization, Funding acquisition, Methodology, Project administration, Resources, Writing – review $\&$
editing.
\textbf{SHB}: Conceptualization, Funding acquisition, Methodology, Project administration, Resources, Supervision, Writing – review $\&$
editing.

\section*{Declaration of competing interest}

The authors declare that they have no known competing financial interests or personal relationships that could have appeared to influence the work reported in this paper.

\section*{Acknowledgments}

The authors acknowledge support from the U.S. Department of Defense, the Army Research Office under Grant No. W911NF-23-10324 (PMs Drs.\ Denise Ford and Robert Martin).
This work used PSC~Bridges2 and NCSA~Delta through allocation PHY210084 (PI Bryngelson) from the Advanced Cyberinfrastructure Coordination Ecosystem: Services $\&$ Support (ACCESS) program~\citep{boerner2023access}, which is supported by National Science Foundation grants $\#$2138259, $\#$2138286, $\#$2138307, $\#$2137603, and $\#$2138296.

\bibliographystyle{bibsty}
\bibliography{references}

\setcounter{figure}{0}
\setcounter{equation}{0}
\renewcommand{\theequation}{{\rm A}.\arabic{equation}}

\appendix

\newpage
\section{Bayesian optimization (BO)}\label{App:BO}

The core of Bayesian Optimization (BO) is to build a surrogate model of the target function using a Gaussian Process (GP) regression and iteratively select points to evaluate based on this model.
The ability of the GP to model a rich distribution over functions depends entirely on the choice of the covariance function. 
After testing different kernels, we chose the Ard Matérn 5/2 kernel~\citep{snoek2012practical}.
The expected improvement criterion is used for the acquisition function with an exploration-exploitation parameter of 0.01. 
This choice of high exploitation is based on the observation that the evaluation of the EIG at a given design does not meaningfully vary when $N_{\text{EIG}}=1000$ samples are used for estimation.
In practice, we leverage the well-developed \texttt{Matlab} function \texttt{fitrgp} for GPR, along with its \texttt{OptimizeHyperparameters} feature. 
Based on the available evaluations, this function optimizes normalization-related hyperparameters, such as length scales and variances, and decides whether to standardize the data.
The ease of implementation of this algorithm makes it an attractive choice for our framework, enabling efficient optimization of the design while maintaining flexibility in hyperparameter tuning.

\newpage
\section{Ensemble-based four-dimensional variational method (En4D-Var)}\label{App:En4dVar}


The En4D-Var filter can be broken down into a forecast and an analysis step.
Given the initial $N_{\text{En}}$ ensembles 
\begin{align}
    \btQ_0 = \mqty[ \btQ_0^{(1)} & \cdots & \btQ_{0}^{(N_{\text{En}})}],
\end{align}
 the states can be propagated in time using \cref{IMR_det} and the corresponding ensemble bubble radii at time step $k$ can be represented as 
 \begin{align}
     \btY_k = \mqty[ {R^*_k}^{(1)} & \cdots &  {R^*_k}^{(N_\text{En})}].
 \end{align}
For a given observed data set, $\vb*{Y}^{\mathrm{D}}$, and a data assimilation window size, $N_t$, the cost function of En4D-Var is
\begin{align}\label{cost_q}
    J(\vb*{Q}_0) = \frac{1}{2 N_t} \sum_{k=1}^{N_t} \left\lVert\vb*{Y}_k^{\mathrm{D}}-\vb*{Y}_k(\vb*{Q}_0)\right\rVert^2_{\vb*{P}_k} 
    + \frac{1}{2}\left\lVert\vb*{Q}_0-\left<\tilde{\vb*{Q}}_0\right>\right\rVert^2_{\vb*{C}_0}.
\end{align}
The norms for the input and output spaces are
\begin{align}
    \|\vb*{Y}_k\|^2_{\vb*{P}_k} \equiv \vb*{Y}_k^\top \vb*{P}_k^{-1}\vb*{Y}_k
    \quad \text{and} \quad  
    \|\vb*{Q}_0\|^2_{\vb*{C}_0} \equiv \vb*{Q}_0^\top \vb*{C}_0^{-1}\vb*{Q}_0 ,
\end{align}
where $\vb*{P}_k$ is the measurement noise covariance matrix at time step $k$, and $\vb*{C}_0=\btQ'_0 {\btQ'_0}^\top$ is the initial ensemble covariance defined using the initial state perturbation matrix, 
\begin{align}
    \btQ'_0 = \frac{1}{\sqrt{N_{\text{En}}-1}}
    \mqty[ \btQ_0^{(1)}-\left<\btQ_0\right>  & \cdots & \btQ_{0}^{(N_{\text{En}})}-\left<\btQ_0\right>],
\end{align}
where $\left<\cdot\right>$ is the ensemble average. 
The optimization for the cost function in \cref{cost_q} is carried out using the form $\vb*{Q}_0  = \tilde{\vb*{Q}}_0+\btQ'_0\cdot\vb*{w}$ to restrict the solution to the subspace spanned by the scaled perturbation matrix around the initial ensembles using the correction coefficient $\vb*{w}$. 
This process is equivalent to finding the minimizer 
\begin{align}
    \vb*{w}_{\text{opt}} = \argmin_{\vb*{w}} J_w(\vb*{w})
\end{align}
for the cost function 
\begin{align}
    J_w(\vb*{w}) =  \frac{1}{2N_t} \sum_{k=1}^{N_t} \left\lVert\vb*{Y}_k^{\mathrm{D}}-\left<\btY_k\right>-\left<\btY_k'\cdot \vb*{w}\right>\right\rVert^2_{\vb*{P}_k} + \frac{1}{2}\vb*{w}^\top\vb*{w},
\end{align}
where the scaled output perturbation matrix takes the form of
\begin{align}
    \btY'_k = \frac{1}{\sqrt{N_{\text{En}}-1}}
    \mqty[ \btY_k^{(1)}-\left<\btY_k\right>  & \cdots & \btY_{k}^{(N_{\text{En}})}-\left<\btY_k\right>].
\end{align}

In practice, we follow \citet{bocquet2013iterative} to seek the optimal correction coefficient $\vb*{w}_\text{opt}$ iteratively using a Gauss--Newton method,
\begin{align}
    \vb*{w}_{i+1}=\vb*{w}_i-\vb*{H}_i^{-1}  \nabla J_i(\vb*{w}_i),
\end{align}
where $i<N_\text{iter}$ is the iteration index, and $\vb*{H}$ and ${\nabla J}$ represent approximations of the Hessian and gradient of $J$.
They can be found with
\begin{align}
    \vb*{H}_i & = ({N_{\text{En}}-1})\vb*{I} 
    +\frac{1}{N_t}\sum_{k=1}^{N_t}  {\btY'_k}^\top\vb*{P}_k^{-1}{\btY'_k}, \\
    {\nabla J}_i & = -\frac{1}{N_t}\sum_{k=1}^{N_t}  {\btY'_k}^\top\vb*{P}_k^{-1}\left(\vb*{Y}_k^{\mathrm{D}} - \left<\btY_k\right>\right)  
    + ({N_{\text{En}}}-1)\vb*{w}_{i}.
\end{align}

\begin{figure}
    \centering
    \includegraphics[scale=1]{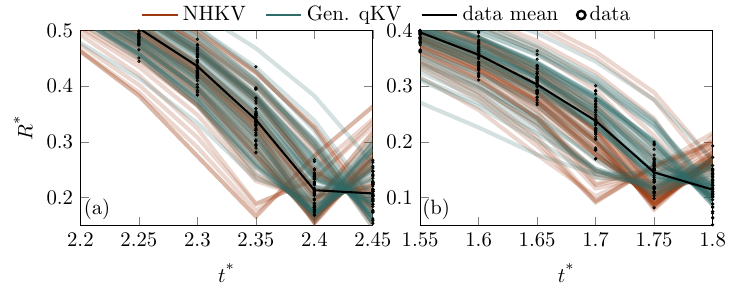}
    \caption{
    Magnified regions of \cref{f:model selection} near the second bubble collapse: (a) \cref{f:model selection}~(a, c) with $p(\text{NHKV})=0.413$ and  $p(\text{Gen\, qKV})=0.587$; (b) \cref{f:model selection}~(b, d) with $p(\text{NHKV})=0.001$ and  $p(\text{Gen\, qKV})=0.999$.
    }
    \label{f:model selection_app}
\end{figure}

By combining the En4D-Var method with the subsequent marginal likelihood calculation, we establish a framework for model inference based on the data.
\Cref{f:model selection_app} shows the zoom-in regions of \cref{f:model selection} near the second bubble collapse, where the differences between the two models are most pronounced.
In \cref{f:model selection_app}~(a), the Gen.\ qKV model performs slightly better than the NHKV model, as reflected in their similar model probabilities.
In \cref{f:model selection_app}~(b), the NHKV model fails to capture the second collapse, whereas some instances of the Gen.\ qKV model do.
This discrepancy, reflected in their model probabilities, results in more confidence in the Gen.\ qKV model.

\begin{figure}[ht]
    \centering
    \includegraphics[scale=1]{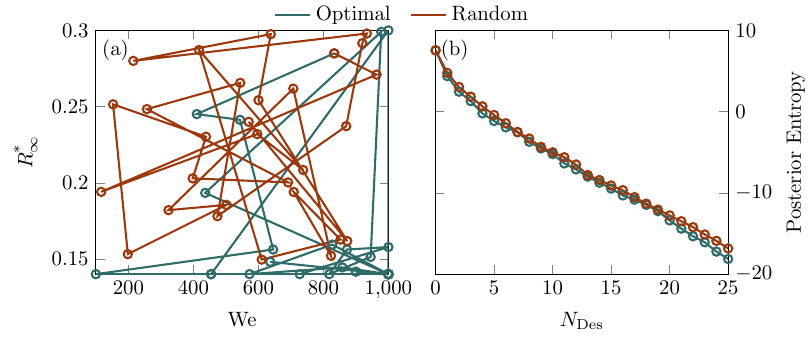}
    \caption{
        Sequential BOED outputs for exploring the stiff PA in~\cref{S:stiff PA} over the design number $N_\mathrm{Des}$: (a) trajectory of the design parameter; (b) posterior entropy.
    }
    \label{f:sequential_results}
\end{figure}

The trajectory of the design parameter used to explore the stiff PA in~\cref{S:stiff PA} over the design number $N_\mathrm{Des}$ is shown in~\cref{f:sequential_results}~(a).
The optimal design tends to explore regions with a larger maximum bubble radius but a smaller equilibrium radius, resulting in a larger stretch ratio.
The variance of the multivariate variable, $\Sigma$, is shown in~\cref{f:sequential_results}~(b) in terms of the posterior entropy, defined as $\log{|\vb*{\Sigma}|}/2+3(1+\log{2\pi})/2$.
A reduction in the variances to levels considered negligible can be observed.

\end{document}